\documentclass[cmp]{svjour}
\usepackage{graphicx}
\usepackage{mathptmx}
\usepackage{amsmath}
\usepackage{amssymb}
\renewcommand{\H}{\text{H}}
\renewcommand{\S}{\text{S}}
\allowdisplaybreaks
\begin{document}
\title{Distribution Functionals for Hard Particles in N Dimensions}
\author{Stephan Korden}
\institute{Institute of Technical Thermodynamics, RWTH Aachen
University, Schinkelstra\ss e 8, 52062 Aachen, Germany
\email{stephan.korden@rwth-aachen.de}}
\date{Version: \today}
\maketitle

\begin{abstract}
The current article completes our investigation of the hard-particle interaction by determining their distribution 
functionals. Beginning with a short review of the perturbation expansion of the free-energy functional, we derive two
representations of the correlation functionals in rooted and unrooted Mayer diagrams, which are related by a functional 
derivative. This map allows to transfer the mathematical methods, developed previously for unrooted diagrams, to the 
current representation in rooted graphs. Translating then the Mayer to Ree-Hoover diagrams and determining their 
automorphism groups, yields the generic functional for all r-particle distributions. From this we derive the examples 
of 2- and 3-particle correlations up to four intersection centers and show that already the leading order reproduces 
the Wertheim, Thiele, Baxter solution for the contact probability of spheres. Another calculation shows the failure of 
the Kirkwood superposition approximation for any r-particle correlation.
\end{abstract}
\section{Introduction}\label{sec:intro}
Density functional theory (DFT) for classical particles introduces direct correlation and distribution functionals. 
These two classes form the building blocks not only for the grand canonical potential and its perturbative expansion but 
also for the Enskog integrals in non-equilibrium thermodynamics and the background potential in quantum mechanical 
calculations \cite{evans-dft,evans-rev,mcdonald}. But contrary to the correlation functions, which can be derived from 
molecular dynamic or Monte Carlo simulations, the correlation functionals are not readily available by numerical 
methods, which explains why so much less is known about this essential part of classical DFT.

The two types of correlations derive from the free energy and the grand canonical potential respectively and are
therefore canonically conjugate variables of their Legendre-dual potentials. This connection allows, at least in 
principal, to write any thermodynamic expression in only one class of correlations and to use integral equations to 
substitute the other \cite{mcdonald}. But solving these identities is a difficult task as the Ornstein-Zernike equation 
shows, whose only known analytic solution has been obtained by Wertheim, Thiele, and Baxter for the interaction of hard 
spheres \cite{baxter,thiele,wertheim-py1,wertheim-py2}. But despite its accuracy for low to medium packing fractions 
and its extension to the mean spherical approximation \cite{mcdonald}, its solution is of limited use for more general 
geometries as the underlying mathematical methods do not generalize to non-spherical interactions. Another and even more 
fundamental restriction for the use of integral equations is their dependence on the correlation functions instead of 
the functionals, which excludes their application for inhomogeneous fluids.

To gain some insight into the structure of the correlation functionals, the current article follows a different 
strategy and starts from the virial expansion in rooted Mayer diagrams \cite{korden-2,korden-4}. For a general 
potential this ansatz with its infinite number of integrals is inaccessible. But for the hard-particle interaction it 
has recently been shown that the free-energy functional expands in a fast converging series of intersection kernels, 
generalizing previous results from Rosenfeld's fundamental measure theory 
\cite{rosenfeld-structure,rosenfeld1,rosenfeld2,rosenfeld-gauss2,tarazona-rosenfeld}, the transformation from Mayer to 
Ree-Hoover diagrams \cite{rh-1,rh-3,rh-2}, Wertheim's derivation of the third virial coefficient 
\cite{wertheim-1,wertheim-2,wertheim-3,wertheim-4}, and the Blaschke, Santalo, Chern equation from integral geometry 
\cite{blaschke,chern-1,chern-2,chern-3,santalo-book}. Its leading contribution coincides with the Rosenfeld functional, 
which has been shown to accurately predict the phase diagram of spheres and polyhedrons from low to medium packing 
fractions 
\cite{schmidt-mixtures,mecke-fmt,mecke-fmt2,marechal-1,marechal-3,marechal-2,cros-ros-1,cros-ros-3,schmidt-dft}. It is 
therefore a natural step to generalize these methods to the more general correlations functionals.

As the direct correlation functionals derive from the free energy, they will be ignored in the following discussion. The 
same applies to the intersection kernels, whose derivation is independent of the virial expansion. This leaves us to 
transfer the diagrammatic methods from the unrooted Ree-Hoover graphs to its rooted form, which will be done in 
Sec.~\ref{sec:corr}, where the generic distribution functional is derived. Examples will then be presented in 
Sec.~\ref{sec:examples}, where the 2- and 3-particle distributions for up to four intersection centers are given. It is 
then shown that already the leading term of the contact probability between equally sized spheres agrees with the 
solution from Wertheim, Thiele, and Baxter. We conclude with a final comment on the applicability of the Kirkwood 
superposition approximation.
\section{The Generic Distribution Functional for Hard Particles}\label{sec:corr}
The main difference between the free-energy and the distribution functionals is their respective virial expansion in 
Mayer clusters. Once this is known, it is a mere technicality to identify their intersection classes and to write the 
functional in intersection kernels. In the following, we will therefore first derive the Mayer representation of the 
r-particle distribution functionals $\rho_{i_1\ldots i_r}$, or more conveniently of their normalized form $g_{i_1\ldots 
i_r}$, translate them into Ree-Hoover diagrams, from which follows the generic correlation functional.

A convenient starting point for the derivation of the Mayer representation is the perturbation expansion of the grand 
canonical potential \cite{mcdonald}. Introducing the hard $\phi^\H$ and soft $\phi^\S$ contributions of the interaction 
potential $\phi_{ij} = \phi^\H_{ij} + \phi^\S_{ij}$ results in a corresponding splitting of the Boltzmann functions 
\begin{equation}\label{F-coupl}
e_{ij} = e_{ij}^\H +  e_{ij}^\H\, f_{ij}^\S = e_{ij}^\H +  \lambda F_{ij} 
\quad \text{for} \quad \lambda = 1\,,
\end{equation}
where we introduced the auxiliary variable $\lambda$ for counting the number of $F$-terms. Next observe that the 
partition function of $N$ particles is a fully e-bonded cluster integral of $N$ labeled nodes $\Gamma_N(e)$ and that 
the expansion of the product $\prod(e_{ij}^\H+\lambda F_{ij})$ yields a sum of products, with a subset of e-bonds 
replaced by F-bonds. Using the invariance of the partition function under relabeling of particle numbers, it is always 
possible to define a unique, labeled subgraph $\Gamma_{r,k}(e^\H, F)$ of $r$ nodes and counting index $k$, such that 
each node is linked to at least one F-bond. The Taylor expansion of the partition function in $\lambda$ can now be 
written as a functional derivative of cluster diagrams
\begin{equation}\label{d-lambda}
D_\lambda = \sum_{r=2}^\infty\sum_{k=1}^{|\Gamma_{r}|}\;\frac{\lambda^{[\Gamma_{r,k}|}}{|\Gamma_{r,k}|!}\; 
\frac{\sigma_{r,k}}{r!}\; \Gamma_{r,k}(e^\H, F) \frac{\delta }{\delta \Gamma_r(e)}\Big|_{\lambda=0}\;,
\end{equation}
where $|\Gamma_{r,k}|$ denotes the number of F-bonds, $|\Gamma_r|$ the total number of $\Gamma_{r,k}$ diagrams, and 
$\sigma_{r,k}$ the number of inequivalent particle labelings.

The Taylor expansion of the partition function $\Xi_\lambda$ up to second order includes the three leading diagrams, 
whose functional derivatives
\begin{align}\label{z-exp}
& \Xi_\lambda = \sum_{N=0}^\infty \frac{z^N}{N!} \int \prod_{i,j=1}^N(e_{ij}^\H + \lambda F_{ij})\, d\gamma_{i_1\ldots 
i_N} = \sum_{N=0}^\infty \frac{z^N}{N!} Z_N(0)\big[1 + D_\lambda \frac{Z_N(\lambda)}{Z_N(0)} \big] \\
&= \Xi_0\big[1+ \lambda\int \frac{1}{2}\rho_{i_1i_2}^\H f_{i_1i_2}^\S\,d\gamma_{i_1i_2} + \frac{\lambda^2}{2}\int
(\rho_{i_1i_2i_3}^\H f_{i_1i_2}^\S f_{i_2i_3}^\S + \frac{1}{4}\rho_{i_1i_2i_3i_4}^\H f_{i_1i_2}^\S 
f_{i_3i_4}^\S)\,d\gamma_{i_1i_2i_3i_4}\nonumber\\[0.2em]
& \hspace{3.25em} + \mathcal{O}(\lambda^3)\big]\nonumber
\end{align}
introduces the grand-canonical r-particle distribution functionals $\rho^\H_{i_1\ldots i_r}$. Expanding its logarithm 
and setting $\lambda=1$ reproduces the well known perturbation expansion of the grand-canonical potential \cite{mcdonald}
\begin{equation}\label{gc-pert}
\begin{split}
\beta \Omega = \beta \Omega_\H & - \frac{1}{2}\int \rho_{i_1i_2}^\H f_{i_1i_2}^\S\,d\gamma_{i_1i_2} - 
\frac{1}{2}\int \rho_{i_1i_2i_3}^\H f_{i_1i_2}^\S f_{i_2i_3}^\S\,d\gamma_{i_1i_2i_3}\\
& - \frac{1}{8}\int(\rho_{i_1i_2i_3i_4}^\H - \rho_{i_1i_2}^\H \rho_{i_3i_4}^\H)f_{i_1i_2}^\S 
f_{i_3i_4}^\S \, d\gamma_{i_1i_2i_3i_4} - \ldots
\end{split}
\end{equation}
where an implicit sum over paired indices is understood. Higher order corrections are determined likewise by successive 
insertion of further F-bonds into the cluster diagrams.

The operator $D_\lambda$ provides a compact notation for the correlation functionals, where each diagram 
$\Gamma_{r,k}(e^\H, F)$ corresponds to exactly one $\rho_{i_1\ldots i_r}^H$. To change the representation into Mayer 
diagrams, one only has to insert $F=e^\H f^\S$, $e^\H = f^\H + 1$ and to expand the graph in the virial series
\begin{equation}\label{vir-t}
\Gamma_{r,k}(f^\H + 1, F) = [f^\S]^{|\Gamma_{r,k}|}\sum_{n\geq r}^\infty \Gamma_{n,k}^{(r)}(f^\H, e^\H)\;,
\end{equation}
which yields a corresponding representation in terms of r-rooted Mayer diagrams $\Gamma_{n,k}^{(r)}$
\begin{equation}\label{gr-vir}
g_r(\vec r_1, \ldots, \vec r_r) = \sum_{n\geq r} \sum_k \frac{\sigma_{n,k}^{(r)}}{(n-r)!} \int 
\Gamma_{n,k}^{(r)}(f, e) \, \rho_{i_{r+1}}\ldots \rho_{i_n}\,d\gamma_{i_{r+1}}\ldots d\gamma_{i_n}\;,
\end{equation}
where we omitted the hard-particle index and introduced the symmetry factor $\sigma_{n,k}^{(r)}$, transforming from 
labeled to unlabeled graphs \cite{riddell,uhlenbeck-ford-1}.

\begin{figure}
\centering
\includegraphics[width=3.0cm,angle=-90]{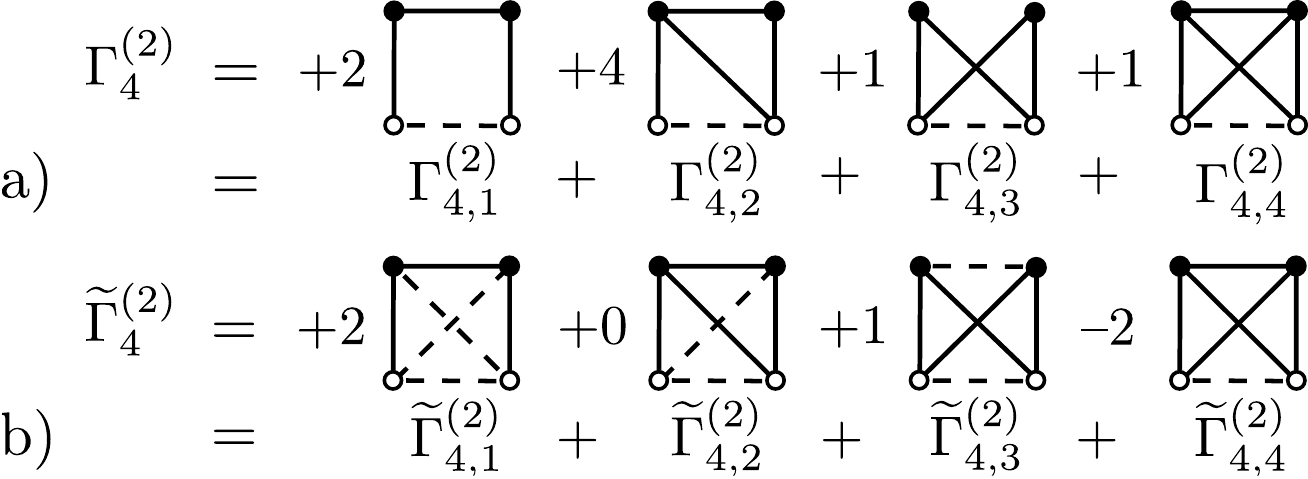}
\caption{The fourth virial order of the 2-particle correlation function in a) Mayer and b) Ree-Hoover diagrams. The 
continuous lines denote f-bonds, while the dashed lines correspond to e-bonds. The rooted points are always mutually 
e-bonded.}
\label{fig:m_rh}
\end{figure}

An alternative representation can be derived by observing that completely e- and f-bonded graphs are uniquely related 
$\Gamma_r(e) \sim \Gamma_r(f)$ by the substitution $e = f + 1$ and ignoring all diagrams of lower order $|\Gamma_{r,k}| 
< |\Gamma_r|$:
\begin{equation}
\Gamma_r(e) = \Gamma_r(f) + \sum_k^{|\Gamma_{r,k}| < |\Gamma_r|} \Gamma_{r,k}(f)\;.
\end{equation}
This relation allows to replace $\Gamma_r(e)$ in (\ref{d-lambda}) by $\Gamma_r(f)$ and to change the representation of 
$D_\lambda$ from the fully connected e-bonded graphs to Mayer diagrams. The resulting operator then applies to the 
virial expansion of $\Omega$, where the functional derivative substitutes any subgraph $\Gamma_r(f) \subset 
\Gamma_{n,k}(f)$ in a Mayer diagram by $\Gamma_r(e)$
\begin{equation}\label{gr-free}
g_r = \Gamma_r(e)\frac{\delta}{\delta \Gamma_r(f)} \Omega\;,
\end{equation}
which reproduces the definition of the normalized r-particle functional. At the level of individual diagrams, this 
operation can also be written as
\begin{equation}\label{r-deriv}
\Gamma_{n,k}^{(r)}(f,e) = \Gamma_r(e)\frac{\delta}{\delta \Gamma_r(f)} \Gamma_{n,k}(f)\;.
\end{equation}
Examples for the 2-rooted diagrams of fourth virial order are shown in Fig.~\ref{fig:m_rh}a).

In the following, we will use the close relationship between the virial expansion (\ref{gr-vir}) and the functional 
derivative (\ref{gr-free}) to transfer the methods previously developed for the free energy to derive the r-particle 
correlation functionals. This approach is divided into two steps: the first one translates the Mayer into intersection 
diagrams, while the second determines their intersection probabilities. Let us first summarize the central ideas and 
notations from \cite{korden-4}:

The approximation method for the free-energy functional uses an expansion in the number of intersection centers, 
whereas the virial series is an expansion in increasing powers of the particle density. Both are uniquely related by 
Ree-Hoover (RH) diagrams $\widetilde\Gamma_{n,k}$, which derive from Mayer graphs $\Gamma_{n,k}$ by inserting 
$1=e_{ij}-f_{ij}$ for each pair $i,j$ of nodes not bonded by an f-function \cite{rh-1,rh-3,rh-2}. Their respective 
cluster integrals are related by the ``star-content'' $a_{n,k}$ as introduced in \cite{rh-1}:
\begin{equation}\label{a_0}
\Gamma_n = \sum_{k'} \Gamma_{n,k'} = \sum_k a_{n,k}\; \widetilde\Gamma_{n,k}\;,
\end{equation}
which satisfies the recursion relations 
\begin{equation}\label{pi_0}
\pi^{-1}(\Gamma_n) = (n-2)\;\Gamma_{n-1}\;,\quad a_{n,k} = (-1)^{n-1} (n-2)\, a_{n-1,k'}
\end{equation}
under removal of a node and its associated f-bonds from a labeled diagram $\pi^{-1}:\Gamma_{n,k} \to \Gamma_{n-1,k'}$. 
This map can be reversed for a nontrivial RH-graph by adding a node and bonding it by f-functions to all previous 
vertices $\pi: \widetilde\Gamma_{n-1,k} \to \widetilde\Gamma_{n,k'}$. For any lowest element $\pi^{-1}(\widetilde 
\Gamma_{n_0,k})=0$, this defines a unique RH-class
\begin{equation}\label{class_0}
\widetilde\Lambda_{n_0,k} = \bigcup_{m=0}^\infty \pi^m(\widetilde\Gamma_{n_0,k})\;,
\end{equation}
whose intersection networks can be contracted into a common pattern with an equal number of intersection centers. Each 
intersection diagram therefore belongs to a unique RH-class, whose elements can be summed up into a generic functional, 
weighted by the numerical prefactors of the virial expansion
\begin{equation}\label{aut_0}
\widetilde\sigma_{n,k} = -\frac{1}{n!}\sigma_{n,k}\,a_{n,k}\;,\quad
\sigma_{n,k} = \frac{|S_n|}{|\text{Aut}(\Gamma_{n,k})|}\;,
\end{equation}
where the symmetry factor $\sigma_{n,k}$ counts the number of inequivalent labelings, determined by the coset of
permutations $S_n$ and the automorphism group $\text{Aut}(\Gamma_{n,k})$. 

The approximate free-energy functional derives from the sum over all contracted intersection diagrams $\gamma_a^I$ of a 
RH-class, where the particle domains $D_{i_1}, \ldots, D_{i_m} \subset \mathbb{R}^N$ of index $I = (i_1, \ldots, i_m)$ 
intersect in the center $\vec r_a \in D_{i_1} \cap \ldots \cap D_{i_m}$. Its functional then factorizes into a 
convolute of integral kernels $K$, satisfying the rules
\begin{equation}\label{rule_1}
\begin{split}
&K(\gamma_a^{\,I_1} + \gamma_b^{\,I_2}) = K(\gamma_a^{\,I_1}) + K(\gamma_b^{\,I_2})\;,\qquad
K(\gamma_a^{\,I_1}\gamma_a^{\,I_2}) = K(\gamma_a^{\,I_1})\, K(\gamma_b^{\,I_2})\;,\\
&\hspace{6.8em} K(\gamma_a^{\,I} [e\ldots e]) = K(\gamma_a^{\,I}) [e\ldots e]
\end{split}
\end{equation}
for the intersection networks $\gamma_a^{\,I_1}$, $\gamma_b^{\,I_2}$, and the product of Boltzmann functions $[e\ldots 
e]$. 

The integral kernel for $N$-dimensional particles in the $N$-dimensional, Euclidean space $\mathbb{R}^N$ is a 
combination of $N+1$ weight functions $w_k^{i_1\ldots i_k}$, determined by the functional derivative with respect to 
the weight function of the particle domain $w_0^i$
\begin{equation}\label{rule_2}
K(\gamma_a^{i_1\ldots i_m}) = \mathcal{D}_a \;w_0^{i_1}(\vec r_{ai_1})\ldots w_0^{i_m}(\vec r_{ai_m})\;,
\end{equation}
where the derivative at intersection point $\vec r_{ai} = \vec r_a - \vec r_i$ is defined by
\begin{equation}\label{rule_3}
\mathcal{D}_a = \sum_{k=1}^N \sum_{(i)} \frac{1}{k!}w_k^{i_1\ldots i_k}(\vec r_{ai_1}, \ldots, \vec r_{ai_k}) 
\frac{\delta^k}{\delta w_0^{i_1}(\vec r_{ai_1})\ldots \delta w_0^{i_k}(\vec r_{ai_k})}\;.
\end{equation}

These definitions provide a set of rules, which are sufficient to derive approximations of the free-energy functional 
for any number of intersection centers. And, as will be shown in the following, they also apply to the correlation 
functionals, replacing star-diagrams by rooted graphs and a corresponding change in their automorphism groups.

The main step in the construction of the correlation functional is again the transformation of the virial series 
(\ref{gr-vir}) from rooted Mayer diagrams to a corresponding set of intersection networks. And in analogy to the 
star-graphs, this transformation requires the intermediate step of inserting $1=e_{ij} -f_{ij}$ for each pair $i,j$ of 
nodes not bonded by f-functions, resulting in a change of the virial series from Mayer to r-rooted RH-diagrams 
$\widetilde \Gamma_{n,k}^{(r)}$ \cite{rh-3,rh-4}. Examples for 2-rooted graphs of the fourth virial order are shown in 
Fig.~\ref{fig:m_rh}b).

Following the conventions of \cite{korden-4}, we define the notation for rooted diagrams:
\begin{definition}
Let $\Gamma_{n,k}^{(r)}$ denote a labeled r-rooted Mayer diagram with $r$ white and $n-r$ black nodes. The rooted 
points are  mutually e-bonded, while the black points are 1-path connected in the subset of f-bonds. 

A black node can be removed by deleting its vertex and all associated f-bonds $\pi^{-1}: \Gamma_{n.k}^{(r)} \to 
\{\Gamma_{n-1,k'}^{(r)}, \Gamma_{n-1,t}^{(r)}\}$, leaving a residual diagram, which is either a new r-rooted Mayer 
graph $\Gamma_{n-1,k'}^{(r)}$ or a sum of disjunct diagrams with articulation points $\Gamma_{n-1,t}^{(r)}$.
\end{definition}
\begin{definition}
Let $\widetilde\Gamma_{n,k}^{(r)}$ denote a labeled r-rooted RH-diagram with $r$ white and $n-r$ black nodes. The 
rooted points are mutually e-bonded, while the black points are 1-path connected in the subset of f-bonds.

A black node without e-bonds can be removed by deleting its vertex and all associated f-bonds $\pi^{-1} : 
\widetilde\Gamma_{n,k}^{(r)} \to \{ \widetilde\Gamma_{n-1,k'}^{(r)}, \; 0\}$, leaving either a new or the trivial 
RH-graph.
\end{definition}

To rewrite the virial integrals (\ref{gr-vir}) in rooted RH-diagrams, observe that the rooting-process (\ref{r-deriv}) 
only exchanges fully f-bonded subdiagrams $\Gamma_r(f)\subseteq \Gamma_{n,k}$ by its corresponding e-bonded graph 
$\Gamma_r(e)$, whereas the Ree-Hoover transformation only operates on pairs of unbonded nodes. The two operations are 
therefore mutually exclusive and commute. From this follows that the same functional derivative (\ref{r-deriv}), which 
transforms star-diagrams into rooted Mayer graphs, also applies to RH-diagrams
\begin{equation}\label{rh-deriv}
\Gamma_r(e)\frac{\delta}{\delta \Gamma_r(f)} \widetilde\Gamma_{n,k} = \widetilde\Gamma_{n,k}^{(r)}\;.
\end{equation}
The previously derived results for unrooted RH-graphs therefore remain valid for its rooted forms.

This connection can be immediately applied to rewrite the transformation between Mayer and RH-diagrams
\begin{equation}
\widetilde\Gamma_{n,k}^{(r)}=\sum_{k'}^{\widetilde\Gamma_{n,k}^{(r)}\subseteq\Gamma_{n,k'}^{(r)}}\Gamma_{n,k'}^{(r)}\;,
\quad \Gamma_{n,k}^{(r)} = \sum_{k'}^{\Gamma_{n,k}^{(r)}\subseteq \widetilde\Gamma_{n,k'}^{(r)}} 
(-1)^{|\Gamma_{n,k}^{(r)}|-|\widetilde\Gamma_{n,k'}^ {(r)}|}\; \widetilde\Gamma_{n,k'}^{(r)}
\end{equation}
and to express the sum over rooted RH-diagrams
\begin{equation}\label{rh-s}
\Gamma_n^{(r)}\; =\; \sum_{k'} \;\Gamma_{n,k'}^{(r)} \;= \;\sum_k a_{n,k}^{(r)}\; \widetilde\Gamma_{n,k}^{(r)}
\end{equation}
in terms of the ``root-content''
\begin{equation}
a_{n,k}^{(r)} = \sum_{k'}^{\Gamma_{n,k'}^{(r)}\subseteq \widetilde\Gamma_{n,k}^{(r)}} (-1)^{|\Gamma_{n,k'}^{(r)}|- 
|\widetilde\Gamma_{n,k}^{(r)}|}\;,
\end{equation}
which satisfies an analogous recursion relation as the star-content (\ref{pi_0}) under removal of a black node. This is 
readily seen by commuting $\pi^{-1}$ with the functional derivative (\ref{r-deriv}) and using (\ref{pi_0})
\begin{equation}\label{lead}
\pi^{-1}(\Gamma_n^{(r)}) = \Gamma_r(e)\frac{\delta}{\delta \Gamma_r(f)}\pi^{-1}(\Gamma_n) = \Gamma_r(e) 
\frac{\delta}{\delta \Gamma_r(f)}(n-2) \,\Gamma_{n-1} = (n-2)\,\Gamma_{n-1}^{(r)}\;,
\end{equation}
with the corresponding relation for RH-diagrams, where the removal of a black node and its $n-1$ f-bonds
\begin{equation}
\begin{split}
\pi^{-1}(\Gamma_n^{(r)}) &= \sum_k \pi^{-1}(a_{n,k}^{(r)}\,\widetilde\Gamma_{n,k}^{(r)})
= (-1)^{n-1}\, \sum_{k'} a_{n,k'}^{(r)}\,\widetilde\Gamma_{n-1,k'}^{(r)}\\
& = (n-2)\,\Gamma_{n-1}^{(r)} = (n-2)\,\sum_{k'} a_{n-1,k'}^{(r)}\,\widetilde\Gamma_{n-1,k'}^{(r)}
\end{split}
\end{equation}
yields the recursion relation
\begin{equation}\label{rec}
a_{n,k}^{(r)} = (-1)^{n-1} (n-2)\; a_{n-1,k}^{(r)}\;.
\end{equation}

The successive application of $\pi^{-1}$ traces each diagram to a unique lowest element $\pi^{-1}(\Gamma_{n_0,k}^{(r)}) 
=0$, which is the first element of the RH-class 
\begin{equation}\label{r-class}
\widetilde\Lambda_{n_0,k}^{(r)} = \bigcup_{m=0}^\infty \pi^m(\widetilde\Gamma_{n_0,k}^{(r)})\;,
\end{equation}
defined by the inverse map $\pi: \widetilde\Gamma_{n-1,k}^{(r)} \to \widetilde\Gamma_{n,k'}^{(r)}$ which attaches a 
node to the previous $n-1$ vertices by f-bonds. The set of rooted RH-diagrams therefore separates into RH-classes, 
whose root-contents can be recursively calculated by (\ref{rec}), proving
\begin{lemma}\label{lem-ar}
The root-content of $\widetilde\Gamma_{n,k}^{(r)} \in \widetilde\Lambda_{n_0,k}^{(r)}$ with lowest element 
$\widetilde\Gamma_{n_0,k}^{(r)}$ is determined by
\begin{equation}\label{star-r}
a_{n,k}^{(r)} = (-1)^{\binom{n}{2}-\binom{n_0}{2}} \;\frac{(n-2)!}{(n_0-2)!}\; a_{n_0,k'}^{(r)}\;.
\end{equation}
\end{lemma}

The last step in rewriting the correlation function (\ref{gr-vir}) in RH-graphs is the transition from labeled to 
unlabeled diagrams. As for the star-diagrams (\ref{aut_0}), the symmetry factor $\sigma_{n,k}^{(r)}$ counts the 
number of inequivalent permutations of particle indices, determined by the coset of the permutation and automorphism 
group of rooted diagrams.
\begin{lemma}
Let $\widetilde\Gamma_{n,k}^{(r)} \in \widetilde\Lambda_{n_0,k}^{(r)}$ denote an element of the RH-class with lowest 
element $\widetilde\Gamma_{n_0,k}^{(r)}$. Its inequivalent labelings of the $r$ white and $n-r$ black nodes are  
permuted by the coset group
\begin{equation}\label{perm-1}
S_r\times S_{n-r}/\text{Aut}(\widetilde\Gamma_{n,k}^{(r)})
\end{equation}
whose automorphism group factorizes into the direct product
\begin{equation}\label{perm-2}
\text{Aut}(\widetilde\Gamma_{n,k}^{(r)}) = S_{n-n_0}\times \text{Aut}(\widetilde\Gamma_{n_0,k}^{(r)})\;.
\end{equation}
\end{lemma}
This is shown as follows: The white and black nodes are labeled independently, resulting in the decoupling of the  
permutation group $S_r\times S_{n-r}$, proving (\ref{perm-1}). Whereas the automorphism group factorizes, because each 
of the $n_0$ nodes is linked to at least one e-bond, while the residual $n-n_0$ nodes are completely f-bonded. Any 
exchange of labels between these two groups therefore results in an inequivalent permutation, leaving the $n-n_0$ 
vertices as an invariant set under relabeling, proving (\ref{perm-2}). 

To rewrite the virial expansion (\ref{gr-vir}) in RH-diagrams, let us define the symmetry factor of RH-integrals
\begin{equation}\label{sym-r}
\widetilde\sigma_{n,k}^{(r)} = \frac{1}{(n-r)!} \sigma_{n,k}^{(r)}\, a_{n,k}^{(r)}\;,
\end{equation}
combining the root-content (\ref{star-r}) and the number of inequivalent labelings
\begin{equation}\label{sig-r}
\sigma_{n,k}^{(r)} = \frac{|S_r\times S_{n-r}|}{|\text{Aut}(\widetilde\Gamma_{n,k}^{(r)})|} = \frac{r!(n-r)!}{(n-n_0)!} 
\frac{1}{|\text{Aut}(\widetilde\Gamma_{n_0,k}^{(r)})|}\;,
\end{equation}
which yields the numerical prefactor of the rooted RH-class: 
\begin{corollary}\label{cor:s}
The symmetry factor of the r-rooted RH-diagram $\widetilde\Gamma_{n,k}^{(r)}\in\widetilde\Lambda_{n_0,k}^{(r)}$ with 
lowest element $\widetilde\Gamma_{n_0,k}^{(r)}$ for $r\geq 2$ is determined by
\begin{equation}\label{r-s}
\widetilde\sigma_{n,k}^{(r)} = (-1)^{\binom{n}{2}-\binom{n_0}{2}} \binom{n-2}{n_0-2} 
\frac{r!\;a_{n_0,k}^{(r)}}{|\text{Aut}(\widetilde\Gamma_{n_0,k}^{(r)})|}\;.
\end{equation}
\end{corollary}
This result follows by inserting (\ref{star-r}), (\ref{sig-r}) into (\ref{sym-r}). 

Up to now, no approximation has been made on the virial expansion (\ref{gr-vir}). The next step is therefore to 
simplify the integrals by restricting the number of intersection centers in which the particles are allowed to overlap. 
A useful observation is the following property of intersection diagrams of a RH-class:
\begin{lemma}\label{lem:net}
The intersection network of the class $\widetilde\Lambda_{n_0,k}^{(r)}$ is defined by its lowest subgraph.
\end{lemma}
The proof begins with the intersection diagram of the lowest element. Any further particle, added by $\pi$, can then be 
chosen to overlap with the previous intersection centers. This shows that a new particle can be added without changing 
their number. Repeated operation with $\pi$ then completes the proof. 

The intersection diagram with the lowest number of intersection centers therefore defines the ``backbone'' diagram for 
the entire RH-class. This is the basic idea for the resummation of RH-diagrams of a given RH-class and the 
approximation transferring the virial series (\ref{gr-vir}) to the generic r-particle correlation functional:
\begin{theorem}
Let $\widetilde\Gamma_{n_0,k}^{(r)}$ denote the lowest element of the RH-class $\widetilde\Lambda_{n_0,k}^{(r)}$. The 
generic r-particle correlation functional of the intersection network $\gamma_{a_1}^{\;I_1} \ldots \gamma_{a_p}^{\;I_p} 
[e\ldots e]$ with particle indices $I \in (i_1,\ldots, i_n)$ is determined by
\begin{equation}\label{r-funct}
\begin{split}
&g_{i_1\ldots i_r}|_p(\vec r_{i_1},\ldots, \vec r_{i_r}| \vec r_{a_1},\ldots, \vec r_{a_p}) = 
\widetilde\Gamma_{r,1}^{(r)}(e) \delta_{n_0,r}\delta_{AI}   + (-1)^{|\widetilde\Gamma_{n_0,k}^{(r)}|} 
\frac{r!\;a_{n_0,k}^{(r)}}{|\text{Aut}(\widetilde\Gamma_{n_0,k}^{(r)})|} \\
&\quad \times \sum_{n\geq n_0} \sum_I \binom{n-2}{n_0-2} 
\int K(\gamma_{a_1}^{\;I_1}\ldots \gamma_{a_p}^{\;I_p} [e\ldots e])\, \rho_{i_{r+1}} 
\ldots \rho_{i_n} \, d\gamma_{i_{r+1}} \ldots d\gamma_{i_n}\;,
\end{split}
\end{equation}
with the notation $\delta_{AI}=\delta(\vec r_{a_1i_1})\ldots \delta(\vec r_{a_ri_r}) \delta(\vec r_{a_{r+1}})\ldots 
\delta(\vec r_{a_p})$ for the product of delta-functions.
\end{theorem}
This result follows from inserting (\ref{r-s}) into the cluster expansion (\ref{gr-vir}) and the cancelation of signs 
due to the identity
\begin{equation}
|\widetilde\Gamma_{n,k}^{(r)}| = |\widetilde\Gamma_{n_0,k}^{(r)}| + \binom{n}{2} - \binom{n_0}{2}\;,
\end{equation}
which leaves an overall constant, depending only on the lowest element of the RH-class and an $n$-dependent binomial 
coefficient. An exception provides the leading diagram $n_0=r$ of each r-correlation functional. Without an f-bond, its 
intersection probability vanishes and therefore has to be included separately. Finally, the delta-functions 
$\delta_{AI}$ have been added in (\ref{r-funct}) to achive a symmetric formulation of the integrals
\begin{align}
[g_{i_1i_2}f_{i_1,i_2}](\vec r_{i_1},\vec r_{i_2})& = \int g_{i_1,i_2}|_3(\vec r_{i_1}, \vec r_{i_2}|\vec r_a, \vec 
r_b, \vec r_c) f_{i_1i_2}(\vec r_{i_1}, \vec r_{i_2})\,d\gamma_a d\gamma_b d\gamma_c \label{rep-i}\\
[g_{i_1i_2}f_{i_1,i_2}](\vec r_a,\vec r_b, \vec r_c) & = \int g_{i_1,i_2}|_3(\vec r_{i_1}, \vec r_{i_2}|\vec r_a, \vec 
r_b, \vec r_c) f_{i_1i_2}(\vec r_{i_1}, \vec r_{i_2})\,d\gamma_{i_1} d\gamma_{i_2} \label{rep-a}
\end{align}
in the particle and intersection coordinates, which proves useful in later applications.
\section{Examples of R-Particle Correlation Functionals}\label{sec:examples}
This last section presents four examples. We begin with the explicit derivation of the 2-particle correlation 
functional with two intersection centers and compare its contact probability with the Ornstein-Zernike solution of 
Wertheim, Baxter, and Thiele. The 2- and 3-particle correlations are then calculated for up to four intersection 
centers and compared to the Kirkwood superposition approximation.

The first f-bonded RH-diagram of the 2-particle correlation functional $g_{i_1i_2}$ is $\widetilde\Gamma_{3,1}^{(2)}$, 
whose two white and one black nodes define a backbone diagram with two intersection centers. All further networks of 
the same RH-class can then be contracted to the pattern
\begin{equation}
\widetilde\Lambda_{2,1}^{(2)}\;:\quad e_{i_1i_2}+e_{i_1i_2}\,\gamma_a^{i_1i_3\ldots i_n}\gamma_b^{i_2i_3\ldots i_n}\;.
\end{equation}
Applying the rules (\ref{rule_1}) and (\ref{rule_2}) for the representation of the intersection kernel and introducing 
the notation $w_a^i = w_0^i(\vec r_{ai})$ for the volume weight, the virial series sums up to the generating function 
\begin{equation}
\begin{split}
&g_{i_1i_2}|_2(\vec r_{i_1}, \vec r_{i_2}| \vec r_a, \vec r_b)\\
&=e_{i_1i_2}\delta_{AI} + \sum_{n\geq 2} \sum_{i_3\ldots i_n} \int 
K(\gamma_a^{i_1i_3\ldots i_n} \gamma_b^{i_2i_3 \ldots i_n}\, e_{i_1i_2})\, \rho_{i_3} \ldots \rho_{i_n} d\gamma_{i_3} 
\ldots d\gamma_{i_n}\\
& = e_{i_1i_2}(\delta_{AI}+ \sum_{n\geq 2}\sum_{i_3\ldots i_n} \mathcal{D}_a\mathcal{D}_b \int w_a^{i_1} w_a^{i_3} 
\ldots w_a^{i_n} w_b^{i_2} w_b^{i_3} \ldots w_b^{i_n}\, \rho_{i_3} \ldots \rho_{i_n} \,d\gamma_{i_3} \ldots 
d\gamma_{i_n})\\
&= e_{i_1i_2}(\delta_{AI} + \mathcal{D}_a\mathcal{D}_b\; w_a^{i_1}w_b^{i_2} \sum_{n\geq 2} x_{ab}^{n-2})\;,
\end{split}
\end{equation}
where we introduced the $x$-variable
\begin{equation}
x_{a_1 \ldots a_p} = \int w_0^i(\vec r_{a_1i})\ldots w_0^i(\vec r_{a_p i})\,\rho_i\, d\gamma_i
\end{equation}
from \cite{korden-4} and used the property of (\ref{rule_3}) that the intersection probability for a single particle is 
zero $\mathcal{D}_a w_a^{i_1}=0$. The final correlation functional has then the analytic form
\begin{equation}\label{g_2_2}
g_{i_1i_2}|_2(\vec r_{i_1}, \vec r_{i_2} | \vec r_a, \vec r_b) = e_{i_1i_2}\Bigl(\delta(\vec r_{ai_1})\delta(\vec 
r_{bi_2})+\mathcal{D}_a\mathcal{D}_b \;\frac{w_a^{i_1}w_b^{i_2}}{1-x_{ab}}\Bigr)\;.
\end{equation}

This result is exact up to the third virial order, but significantly improved by the additional pole at $x_{ab}=1$. To 
illustrate this effect of the resummation process, let us derive the leading order of the contact probability 
$g_{i_1i_2}(|\vec r_{i_1} - \vec r_{i_2}|=D^+)$ for spheres of radius $R$ and diameter $D=2R$. The polynomial expansion 
of the numerator of (\ref{g_2_2}) up to first order in $\rho$ yields three terms  
\begin{align}
&g_{i_1i_2}|_2(\vec r_{i_1}, \vec r_{i_2}) = e_{i_1i_2} \int \Bigl( \delta_{AI}+ \mathcal{D}_a 
\mathcal{D}_b \frac{w_a^{i_1}w_b^{i_2}}{1-x_{ab}}\Bigr)\,d\gamma_a d\gamma_b \nonumber\\
&= e_{i_1i_2}\int \Bigl(\delta_{AI} + \mathcal{D}_a \frac{w_a^{i_1}\int w_a^{i_3} \mathcal{D}_b (w_b^{i_2}w_b^{i_3}) 
\rho_{i_3}\, d\gamma_{i_3} + \ldots }{(1-x_{ab})^2}\Bigr)\,d\gamma_a d\gamma_b \label{g_2_exp}\\
&= \int \frac{e_{i_1i_2}}{(1-x_{ab})^2}\Bigl( \delta_{AI}(1-2x_{ab})+ \int (\mathcal{D}_a 
w_a^{i_1}w_a^{i_3})(\mathcal{D}_b w_b^{i_2} w_b^{i_3}) \rho_{i_3}\, d\gamma_{i_3} + \mathcal{O}(\rho^2)\Bigr) 
\,d\gamma_a d\gamma_b \nonumber\;,
\end{align}
of which two correspond to the second and third virial orders, whereas $-2x_{ab}$ originates from resummation. Their 
respective integrals determine the lens-like volumes of overlapping spheres
\begin{equation}
V_R(r) = \frac{4\pi}{3}R^3\Bigl[1-\frac{3}{4}\frac{r}{R} + \frac{1}{16}\Bigl(\frac{r}{R}\Bigr)^3 \Bigr]\;,\qquad
V_D(r) = \frac{4\pi}{3}D^3\Bigl[1-\frac{3}{4}\frac{r}{D} + \frac{1}{16}\Bigl(\frac{r}{D}\Bigr)^3\Bigr]
\end{equation}
and can be readily evaluated for $|\vec r_{i_1}-\vec r_{i_2}|=D$. Observing that the last term corresponds to the third 
virial integral 
\begin{equation}
\int (\mathcal{D}_a w_a^{i_1}w_a^{i_3})(\mathcal{D}_b w_b^{i_3}w_b^{i_2})\rho_{i_3}d\gamma_{i_3} = \delta_{AI} \int 
f_{i_1i_3}f_{i_3i_2}\rho_{i_3}d\gamma_{i_3} = \delta_{AI} \rho \,V_D(r_{i_1i_2}=D)
\end{equation}
and that the intersection coordinates can be chosen to coincide
\begin{equation}
\int x_{ab}\delta(\vec r_{ai_1})\delta(\vec r_{bi_2})\,d\gamma_a d\gamma_b = \rho V_R(\vec r_{ab}=0) = \eta\;,
\end{equation}
the integral (\ref{g_2_exp}) yields the contact probability as a function of the packing fraction $\eta$:
\begin{equation}\label{wer}
\begin{split}
g_{i_1i_2}(r_{i_1i_2}=D^+) &= \frac{1}{(1-\eta)^2}(1-2\eta + \frac{5}{2}\eta) + \mathcal{O} \Bigl( 
\frac{\eta^2}{(1-\eta)^3} \Bigr)\\
&=\frac{1+\frac{1}{2}\eta}{(1-\eta)^2} + \mathcal{O}\Bigl(\frac{\eta^2}{(1-\eta)^3}\Bigr) \;,
\end{split}
\end{equation}
which agrees with the Carnahan-Starling polynomial to first order in the numerator as well as the Wertheim, Thiele, 
Baxter solution of the Ornstein-Zernike equation \cite{mcdonald,thiele,wertheim-py1,wertheim-py2}. The latter 
observation is especially interesting because its closing condition $c_2(r>D)=0$ corresponds to the 1-center 
approximation of the functional expansion, whereas (\ref{wer}) reflects the 2-center representation of $g_2$. To obtain 
the same accuracy for the distribution functional therefore requires a larger number of intersections centers than for 
its dual direct correlation.

Deriving higher order functionals is now a matter of simple algebra. Here we list the leading orders of the 2-particle 
correlations with up to four intersection centers. The corresponding intersection diagrams for the RH-classes in the 
notation of Fig.~\ref{fig:m_rh}b) are obtained by successive contraction of pair-wise intersection centers
\begin{align}
\widetilde\Lambda_{2,1}^{(2)}\;&:\; e_{i_1i_2}+e_{i_1i_2}\,\gamma_a^{i_1i_3\ldots i_n}\gamma_b^{i_2i_3\ldots i_n} 
\nonumber\\
\widetilde\Lambda_{2,1}^{(2)} + \widetilde\Lambda_{4,1}^{(2)}\;&:\;  e_{i_1i_2}+ e_{i_1i_2} \gamma_a^{i_1i_3} 
\gamma_b^{i_3i_2} + e_{i_1i_2} \gamma_a^{i_1i_3i_5\ldots i_n} \gamma_b^{i_2i_3i_4i_5\ldots i_n}
\gamma_c^{i_1i_4i_5\ldots i_n} \nonumber\\
&\quad  + e_{i_1i_2} e_{i_2i_3} e_{i_1i_4} \gamma_a^{i_1i_3i_5\ldots i_n} \gamma_b^{i_3i_4i_5\ldots i_n} 
\gamma_c^{i_2i_4i_5\ldots i_n}\\
\widetilde\Lambda_{2,1}^{(2)} + \widetilde\Lambda_{4,3}^{(2)}\;&:\;
e_{i_1i_2}+ e_{i_1i_2} \gamma_a^{i_1i_3} \gamma_b^{i_3i_2} + e_{i_1i_2}\,\gamma_{a_1}^{i_1i_3i_5\ldots i_n} 
\gamma_{a_2}^{i_1i_4i_5\ldots i_n}\gamma_{b_1}^{i_2i_3i_5\ldots i_n}\gamma_{b_2}^{i_2i_4i_5\ldots i_n} \nonumber\\
&\quad  + e_{i_1i_2} e_{i_3i_4}\,\gamma_{a_1}^{i_1i_3i_5\ldots i_n} \gamma_{a_2}^{i_1i_4i_5\ldots i_n} 
\gamma_{b_1}^{i_2i_3i_5\ldots i_n} \gamma_{b_2}^{i_2i_4i_5\ldots i_n}\;,\nonumber
\end{align}
with the corresponding 2-particle correlation functionals
\begin{align}
g_{i_1i_2}|_2 & = e_{i_1i_2}\Bigl(\delta_{AI} +\mathcal{D}_a\mathcal{D}_b \;\frac{w_a^{i_1}w_b^{i_2}}{1-x_{ab}}\Bigr) 
\nonumber\\
g_{i_1i_2}|_3 & = e_{i_1i_2}\Bigl(\delta_{AI} + \mathcal{D}_a\mathcal{D}_b w_a^{i_1}w_b^{i_2} x_{ab}
+\mathcal{D}_a\mathcal{D}_b\mathcal{D}_c w_a^{i_1} w_b^{i_2} w_c^{i_1} \frac{x_{ab} x_{bc}}{1-x_{abc}} \nonumber\\
&\hspace{4.7em} - \mathcal{D}_a \mathcal{D}_b \mathcal{D}_c \frac{w_a^{i_1}w_c^{i_2} 
y^{i_1i_2}_{abc}}{(1-x_{abc})^3}\Bigr)\\
g_{i_1i_2}|_4 &= e_{i_1i_2}\Bigl(\delta_{AI} + \mathcal{D}_{a_1}\mathcal{D}_{b_1} w_{a_1}^{i_1}w_{b_1}^{i_2} 
x_{a_1b_1} \nonumber\\
&\hspace{4.7em} +\mathcal{D}_{a_1}  \mathcal{D}_{a_2} \mathcal{D}_{b_1} \mathcal{D}_{b_2} (w_{a_1}^{i_1} w_{a_2}^{i_1})
(w_{b_1}^{i_2} w_{b_2}^{i_2}) \frac{x_{a_1b_1} x_{a_2b_2}}{1-x_{a_1a_2b_1b_2}}\nonumber \\
& \hspace{4.7em} + \frac{1}{2} \mathcal{D}_{a_1}  \mathcal{D}_{a_2} \mathcal{D}_{b_1}  \mathcal{D}_{b_2}
\frac{w_{a_1}^{i_1}w_{a_2}^{i_2}w_{b_1}^{i_1}w_{b_2}^{i_2} y_{a_1a_2b_1b_2}}{(1-x_{a_1a_2b_1b_2})^3} \Bigr)\;,\nonumber
\end{align}
where we introduced the Boltzmann weighted densities
\begin{equation}
\begin{split}
y^{i_1i_2}_{abc} &=  \sum_{i_3,i_4} \int e_{i_1i_4}e_{i_2i_3} w_a^{i_3}w_b^{i_3}\,\rho_{i_3}\, w_b^{i_4} w_c^{i_4}\, 
\rho_{i_4}\, d\gamma_{i_3} d\gamma_{i_4}\\
y_{a_1a_2b_1b_2} &= \sum_{i_3,i_4} \int e_{i_3i_4}w_{a_1}^{i_3} w_{b_1}^{i_3}\,\rho_{i_3} w_{a_2}^{i_4} 
w_{b_2}^{i_4}\,\rho_{i_4}\, d\gamma_{i_3} d\gamma_{i_4}\;.
\end{split}
\end{equation}

As a last example, we present the leading term of the 3-particle correlation functional with 3 intersection centers. 
Its intersection diagram has the form
\begin{equation}
\widetilde\Lambda_{3,1}^{(3)}\;: \hspace{4.5em} e_{i_1i_2}e_{i_2i_3} e_{i_1i_3}
+ e_{i_1i_2}e_{i_2i_3} e_{i_1i_3} \gamma_a^{i_1i_4\ldots i_n} 
\gamma_b^{i_2i_4\ldots i_n} \gamma_c^{i_3i_4\ldots i_n}\;, \hspace{7.9em}
\end{equation}
with the corresponding correlation functional
\begin{equation}\label{3-dis}
g_{i_1i_2i_3}|_3 = e_{i_1i_2}e_{i_2i_3}e_{i_1i_3}\Bigl(\delta_{AI} + \mathcal{D}_a \mathcal{D}_b \mathcal{D}_c 
\frac{w_a^{i_1}w_b^{i_2}w_c^{i_3}}{(1-x_{abc})^2}\Bigr)\;.
\end{equation}
The contact probability is therefore to leading order $g_3\sim (1-\eta)^{-3}$, which shows that the Kirkwood 
approximation $g_{i_1i_2i_3}\approx g_{i_1i_2}g_{i_2i_3} g_{i_3i_1} = [g_2]^3 \sim (1-\eta)^{-6}$ is not applicable 
when all three particles are close together, in accordance with results obtained from computer simulations \cite{kirk}. 
It is not difficult to generalize this result to an arbitrary r-particle correlation function, whose leading term is 
$g_r \sim (1-\eta)^{-r}$, whereas the Kirkwood approximation suggests $[g_2]^{r(r-1)/2} \sim (1-\eta)^{-r(r-1)}$. This 
excludes the superposition approximation as a construction principle for any distribution functional.
\section{Discussion and Conclusion}\label{sec:conclusion}
The current derivation of the distribution functionals completes our investigation of the hard-particle correlations, 
which started with the analysis of Rosenfeld's fundamental measure theory. Together with this previous result, it is 
now possible to calculate all direct and distribution functionals in an expansion of intersection kernels,  
highlighting the hard-particle interaction as the only known potential whose correlations can be derived by analytic 
methods instead of molecular dynamic or Monte Carlo simulations.

Despite their complex structure it is possible to derive several characteristic properties from these functionals. For 
example, we have shown that the Wertheim, Thiele, Baxter solution only includes the third virial integral, whereas the 
2-center approximation also contains contributions from the 4-particle diagram. The expansion of the pair correlation 
to leading order in the intersection kernels therefore applies not only to more general particle geometries than 
Wertheim's solution, but also includes approximations of higher virial terms. We also tested the Kirkwood superposition 
approximation, proving that it fails for any distribution functional, a result that has been previously shown 
numerically only for the 3-particle correlations.

These examples demonstrate the difficulties to find appropriate methods to evaluate the non-local correlation 
functionals. A possible approach is, e.g., to generalize Wertheim's ansatz of expanding the intersection kernels in 
convolutes of r-point densities. But the integrals soon become intractable to solve. The only alternative, which also 
applies to concave geometries, is the triangulation of the surfaces and to calculate the functionals numerically. The 
intersection configurations of the particles can then be generated either by discretization of the imbedding space or 
by Monte Carlo sampling as has been done by Ree and Hoover. But the time-consuming step remains the determination of 
the intersection domains and their Euler densities. Both problems have been solved in the framework of discrete 
geometry and implemented in software libraries for 3-dimensional image manipulation. But the vast number of 
intersection configurations that have to be evaluated while minimizing the grand potential requires further 
approximations to reduce the calculational costs in order to be competitive to molecular dynamic and Monte Carlo 
methods.

The only particle geometry for which the functionals can be evaluated by algebraic relations are spheres. Instead of 
looking for approximations of the functionals, it is therefore more appropriate to replace the geometry of molecules by 
sets of overlapping balls. This approach will be discussed in a forthcoming article, where the grand potential for 
realistic molecules of hard and soft interactions will be derived \cite{korden-5}. 
\begin{acknowledgements}
The author wishes to thank Andr\'{e} Bardow and Kai Leonhard for their support of this work. 
\end{acknowledgements}
\bibliographystyle{spmpsci}
\bibliography{fmt_correlations}
\end{document}